\documentclass{article}

\usepackage{arxiv}

\usepackage[utf8]{inputenc} 
\usepackage[T1]{fontenc}    
\usepackage{url}            
\usepackage{booktabs}       
\usepackage{amsfonts}       
\usepackage{nicefrac}       
\usepackage{microtype}      
\usepackage{lipsum}		
\usepackage{graphicx}
\usepackage{doi}
\usepackage{mathrsfs}
\usepackage{hyperref}       

\usepackage{amsmath,amsfonts,amssymb}
\usepackage{graphicx}
\usepackage{setspace}
\usepackage{tocloft}
\usepackage{commands}
\usepackage[normalem]{ulem}

\title{Anisoplanatic Optical Turbulence Simulation for Near-Continuous $C_n^2$ Profiles without Wave Propagation}

\author{ \href{https://orcid.org/0000-0001-9528-0102}{\includegraphics[scale=0.06]{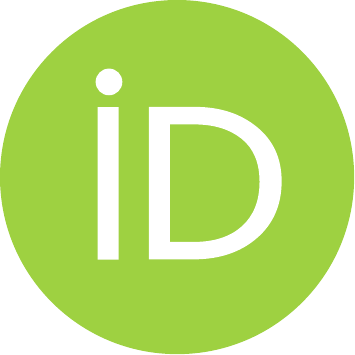}\hspace{1mm}Nicholas Chimitt}
    \\
    School of Electrical and Computer Engineering\\
    Purdue University \\
 	\And
    \href{https://orcid.org/0000-0001-5876-2073}{\includegraphics[scale=0.06]{orcid.pdf}\hspace{1mm}Stanley H.~Chan} \\
	School of Electrical and Computer Engineering\\
	Purdue University\\
}

\date{}

\renewcommand{\shorttitle}{Anisoplanatic Optical Turbulence Simulation for \\Near-Continuous $C_n^2$ Profiles without Wave Propagation}

\hypersetup{
pdftitle={Anisoplanatic Optical Turbulence Simulation for Near-Continuous Cn2 Profiles without Wave Propagation},
pdfauthor={Nicholas Chimitt, Stanley H.~Chan},
pdfkeywords={Zernike polynomials, atmospheric turbulence, phase distortions, simulation},
}

\begin{document}

\maketitle

\begin{abstract}
For the simulation of anisoplanatic optical turbulence, split-step propagation is the gold standard. Within the context of the degradations being limited to phase distortions, one instead may focus on generating the phase realizations directly, a method which has been utilized in previous so-called multi-aperture simulations. Presently, this modality assumes a constant $C_n^2$ profile. This work presents an alternative derivation for Zernike correlations under anisoplanatic conditions. Multi-aperture simulation may easily incorporate these correlations into its framework and achieve a significantly higher degree of accuracy with a minimal increase in time. We additionally use our developed methodology to explain previously reported discrepancies in an empirical implementation of split-step with the analytic tilt correlation. Finally, we outline a major limitation for Zernike-based simulation which still remains.
\end{abstract}

\keywords{Zernike polynomials, atmospheric turbulence, phase distortions, simulation}

\section{Introduction}
Optical wave propagation through a turbulent medium produces a complicated form of wave distortions, affecting both the amplitude and phase. In this work, as in many others we describe here, we focus on the phase distortions generated by the atmosphere. Much of the classical literature focuses on the case of a single point source which has led to a considerable body of work dedicated to the understanding of the statistics of a single wavefront \cite{Tatarski1967, Fried65, Fried66optical, Noll_1976, Fried65_geom}. Among these, Noll \cite{Noll_1976} decomposed the turbulent wavefront distortions by the Zernike polynomials, which is critical to this work. Beyond a single wavefront, the correlation statistics for two wavefront tilts as a function of their separation, otherwise known as angle of arrival correlations \cite{Fried1976_variety, Fried1975_differential, Basu_2015}, have been investigated. These correlations have a long correlation length and contain much of the energy of the distortions. For the sake of anisoplanatic modeling, the angle-of-arrival correlation is an important marker of accuracy.

In this work, we focus on the problem of correlation for \emph{all} Zernike coefficients as a function of their separation in the object plane (or equivalently, their angular separation). Within this space are works \cite{Hu_zernike_corr, Molodij_zernike_corr, Valley_zernike_corr} in which the spatial correlation of Zernike coefficients (or more generally any basis representation) are considered in a variety of situations. Most directly related to our problem is the work of Whiteley et al. \cite{Whiteley98}, choosing to comment on this paper once the major concepts have been suitably described. In contrast to these, we use a comparatively simple approach to arrive at similar results, and demonstrate the applicability of these results to generation of turbulent phase statistics. This has been approximately achieved by Chimitt and Chan \cite{Chimitt2020}, with the limitation of their approach having a simple interpretation within our analysis. With our results, we can match the predicted tilt results of Fried \cite{Fried1976_variety} as a special case of our general result.

Beyond theoretical understanding, our results apply directly to recently proposed multi-aperture simulation approaches by Chimitt et al. \cite{Chimitt2020, Chimitt_2022} and Mao et al. \cite{Mao_2021_ICCV}. These methods use Zernike-based generation concepts to simulate anisoplanatic turbulence on an image, though with two significant limitations. One of these is the limitation of only being applicable to a constant $C_n^2$ profile which significantly limits its usage. However, with adoption of the theoretical groundwork described in this work, these Zernike-based approaches can now generate realizations for near-continuous $C_n^2$ profiles in both constant and arbitrarily varying cases. It is important to mention that in spite of the removal of this significant limitation, there still remain some challenges for the utilization of these results for a full-frame image, which we detail towards the end of this paper.

The most common approach towards simulating these effects on an image is the classical method of split-step propagation \cite{RoggemannSimulator, HardieSimulator}. While the split-step modality is more general than the discussed Zernike-based simulation, within the context of phase distortion modeling Zernike-based models are more suitable for large dataset generation which has been highlighted in previous works \cite{Chimitt2020, Mao_2021_ICCV, Chimitt_2022}. Split-step is based upon numerical wave simulation, which is largely neglected by the computer vision/image processing communities due to speed in generation \cite{Mao_2021_ICCV}. By contrast, Zernike-based simulation's lack of wave propagation while maintaining a high degree of accuracy highlights the core benefit of this modality. The cost to be paid for these benefits is (i) the restriction to phase distortions and (ii) a complicated correlation expression -- though, simple to evaluate numerically.

We outline our main contributions as follows:
\begin{enumerate}
    \item \textbf{Alternative derivation of Zernike correlations:} An approach towards deriving the Zernike coefficient correlations is provided. While the derivation has some similarity to Whiteley et al. \cite{Whiteley98}, our approach is comparatively simplistic and additionally allows for more general numerical evaluation techniques. The derivation provided also has a visual interpretation of the correlations;
    \item \textbf{Introduction of a $C_n^2$-slice:} With the expression for Zernike correlations, we discretize and interpret this result to empower Zernike-based methods \cite{Chimitt2020, Mao_2021_ICCV} for varying turbulence profile simulation, leading us to introduce a $C_n^2$-slice;
    \item \textbf{Zernike-based simulation with $C_n^2$-slices:}
    The $C_n^2$-slice concept, facilitated by the general nature of our result, can be used to replace the correlations of previous methods \cite{Chimitt2020, Mao_2021_ICCV, Chimitt_2022}. This leads us to describe various aspects of the state of Zernike-based simulation's accuracy and speed. Additionally, we outline remaining limitations in Zernike-based simulation methodologies.
\end{enumerate}

\section{Background}
We begin with some definitions of the Zernike polynomials using Noll's conventions as well as some additional notations for tracking source positions. After discussing two simulation approaches and their limitations in more detail, we then present the angle-of-arrival correlations described in Fried \cite{Fried1976_variety} using our notation. Afterwards, we turn to a seemingly unrelated problem of the correlation of the same point source viewed by a two \emph{separate} imaging systems. Ultimately, we will use the results of the two aperture correlations to analyze and directly compare with Fried\cite{Fried1976_variety}.

\subsection{Preliminary Definitions and Single Point Source Statistics}
We define the object plane coordinates to be $\vx = (x, y)$, while we use normalized polar coordinates $\vrho = (\rho,\theta)$ for the coordinates within the aperture plane, with $\rho = 1$ on the edge of the aperture. We denote the propagation distance as $L$, with $z$ as the distance \emph{from} the imaging system (thus $z=0$ is in the aperture plane and $z=L$ in the object plane). We further define $D$ and $R$ to be the aperture diameter and radius, respectively. Finally, we adopt Noll's indexing conventions for mapping the radial and angular components to a single index, $(n,m) \to i$.

The Zernike polynomials $\{Z_i\}$ were famously applied to the problem of turbulent phase distortions by Noll \cite{Noll_1976}. Noll chose to define the Zernike polynomials such that
\begin{equation}
    \frac{1}{\pi} \int d\vrho P(\rho) Z_i(\vrho) Z_j(\vrho) = \delta_{ij},
\end{equation}
with $P(\rho) = 1$ when $\rho \leq 1$. This accordingly defines the Zernike polyonmials over a unit circle. The Zernike polynomials can be used to represent the phase component $\phi(\vrho)$ of a wave originating at point $\vx$ that has propagated through a turbulent medium through basis decomposition,
\begin{equation}
    \phi_{\vx}(R\vrho) = \sum_i a_{\vx,i} Z_i(\vrho).
\end{equation}
We add the subscript $\vx$ to track the position of the point source in the object plane. Noll \cite{Noll_1976} investigates the correlation of two Zernike coefficients for a single point source,
\begin{equation}
    \E [a_{\vx,i} a_{\vx,j}] = \frac{1}{\pi^2} \iint d\vrho d\vrho' P(\rho) P(\rho')  Z_i(\vrho) Z_j(\vrho') \E[\phi_{\vx}(R\vrho) \phi_{\vx}(R\vrho')].
    \label{eq: Nolls_problem}
\end{equation}
This covariance may be re-written using the phase structure function for a spherical wave as
\begin{equation}
    \calD(R\vrho - R\vrho') = 2.91 k^2 \int_0^L dz C_n^2(z) \left|R(\vrho - \vrho')\left(\frac{L - z}{L}\right)\right|^{5/3}.
    \label{eq: struct_fun_sphere}
\end{equation}
We may write the structure function as $\calD(R\vrho - R\vrho') = \E[(\phi_\vx(R\vrho) - \phi_\vx(R\vrho))^2]$, which is independent of position $\vx$. Combined with $\E[\phi_\vx (R\vrho)] = \vec{0}$, we may substitute this in \eqref{eq: Nolls_problem}, giving us
\begin{equation}
    \E [a_{\vx,i} a_{\vx,j}] = \frac{-2.91 k^2}{2\pi^2} \int dz C_n^2(z) \iint d\vrho d\vrho' P(\rho) P(\rho') Z_i(\vrho) Z_j(\vrho') \left|R(\vrho - \vrho')\left(\frac{L - z}{L}\right)\right|^{5/3}.
\end{equation}
The evaluation of this integral is given in the appendix of Noll's paper\cite{Noll_1976}. This result allows for the description and simulation of a single wavefront distorted by atmospheric turbulence using the Zernike polynomials; this result, however, has no direct implication for the statistics beyond a single wavefront or point source. For anisoplanatic simulation based upon the Zernike polynomials we will need a more general expression.

\subsection{Simulation Approaches: Summary and Current Limitations}
There exist a handful of methods to simulate atmospheric turbulence in the literature, with the two main focuses of this work being the split-step \cite{RoggemannSimulator, HardieSimulator, SchmidtTurbBook} and multi-aperture \cite{Chimitt2020, Mao_2021_ICCV, Chimitt_2022} simulations. These methods operate from two fundamentally different perspectives, though split-step is currently the more theoretically justified approach. There is a gap in accuracy between the two methods due to inherent assumptions that were made to simplify the analysis within the multi-aperture method. This work, in part, attempts to minimize this gap.

\subsubsection{Split-Step Simulation}
\begin{figure}
    \centering
        \includegraphics[width=0.95\linewidth]{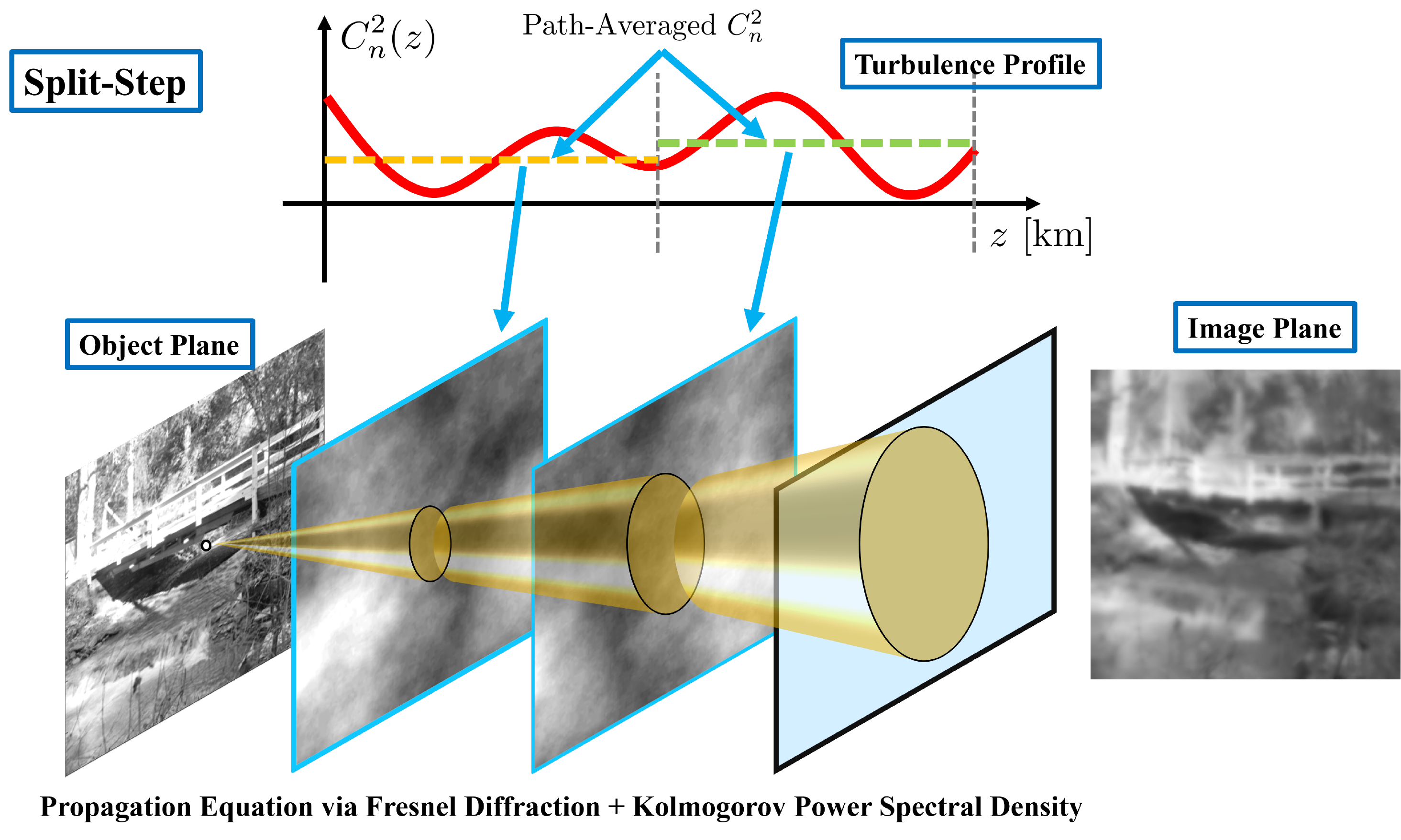}
    \caption{A visualization of split-step propagation. A point (optionally a grid of points) is propagated through a series of phase screens by numerical wave propagation. The result is a collection of phase realizations for each point propagated, which can then be used to form PSFs.}
    \label{fig: split_step_visualized}
\end{figure}

The traditional approach to generating optical phase statistics is that of split-step simulation. Split-step models the forward process of nature directly, making it considerably accurate and intuitive. A split-step simulation can be described by three main steps:
\begin{enumerate}
    \item \textbf{Generate phase screens:} First, discrete phase screens representing the turbulent distortions along the path of propagation are generated. These are typically of a small order, with \cite{HardieSimulator} effectively using 9 for their simulations;
    \item \textbf{Numerical wave propagation:} A source field is generated and propagated via evaluation of the Fresnel integral \cite{SchmidtTurbBook} to the first phase screen. The phase is imparted into the wave, which is then propagated to the next phase screen. The process is then repeated until landing upon the aperture;
    \item \textbf{Image generation:} A point spread function (PSF) is then formed by the incident wave, which can then be applied to the source object. Steps 2 and 3 are then repeated for every point on the object.
\end{enumerate}
We include the final step for completeness, though the core of the split-step simulation is in the phase screen generation and numerical wave propagation. Additionally, sub-sampling of the object plane is typically performed to reduce the simulation time. We give a visualization of the split-step method in \fref{fig: split_step_visualized}. Split-step gains the ability to generate anisoplanatic samples with minimal theoretical effort; the phase screens can be made large enough so that every point in the object plane will pass through the large phase screens. By virtue of the sharing of phase screen components during the wave propagation, as in nature, the phase realizations at the aperture will be correlated according to the theory.

Split-step has a few limitations. From the perspective of generating training data, the most notable is the speed at which the method is able to generate samples. This limitation arises from both (i) the generation of multiple phase screens; (ii) numerical wave propagation. For a complicated $C_n^2$ profile with sudden peaks and valleys, split-step will require a large amount of phase screens to numerically propagate through, a point we shall elaborate on later in this work. This generation and propagation step will need to be done \emph{per realization}. For the usage of split-step in generating data for the purposes of numerical analysis, this will take a potentially infeasible amount of time. Additionally, the number of phase screens have an upper limit by the necessity of their independence.

\subsubsection{Zernike-Based Simulation}
\begin{figure}
    \centering
        \includegraphics[width=0.95\linewidth]{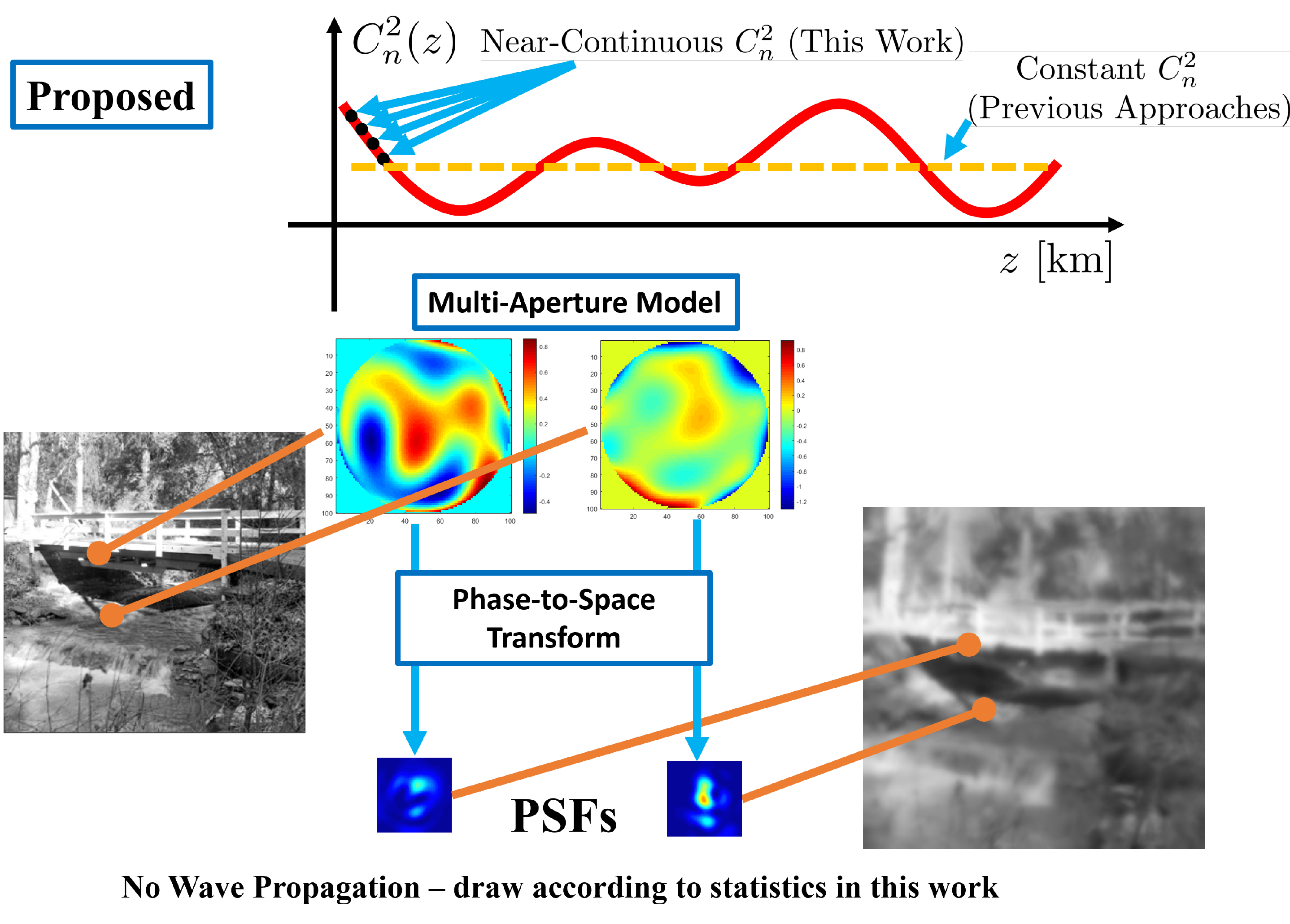}
    \caption{A visualization of the multi-aperture simulation. For each pixel in an image, a Zernike vector is generated, which weights the Zernike polynomials to create a phase realization per-pixel. These phase realizations may then be used to form PSFs.}
    \label{fig: multi_aperture_visualized}
\end{figure}

Zernike-based simulation is fundamentally different than split-step propagation, it does not directly simulate the wave propagation process. Instead, it pulls statistics directly at the aperture plane for each pixel in the image. Therefore, it is limited by the theoretical understanding of spatial statistics of the Zernike coefficients (or alternatively some other basis representation). Previous multi-aperture simulations differ from other Zernike-based simulations such as \cite{roddier_zernikesim} by virtue of their ability to approximately model anisoplanatism. The multi-aperture variety of Zernike-based simulations can be described by two major steps:
\begin{enumerate}
    \item \textbf{Generate Zernike coefficients:} A Zernike coefficient vector is formed for a subset of pixels in the image. This uses Noll's results as a starting point, with some additional modifications proposed by \cite{Chimitt2020};
    \item \textbf{Image Generation:} A PSF is then formed from the Zernike representation (either analytically or via the numerical approach of \cite{Mao_2021_ICCV}) for every pixel in the image, and is then applied to the image.
\end{enumerate}
The core of the simulation rests in the generation of the Zernike coefficients. Drawing samples of the Zernike vectors, correlated both spatially and inter-modally, is the primary focus of this type of simulation of which we give a visualization in \fref{fig: multi_aperture_visualized}.  This also highlights the main reason for the dramatic improvement in speed over split-step: With knowledge of the correlations, there is no need to numerically propagate a wave. The trade-off for Zernike-based simulation is that there must be additional efforts in ensuring the spatial correlations are generated according to theory. Additionally, it is important to note there is not straightforward path towards a principled incorporation of amplitude effects. Therefore, split-step outperforms Zernike-based simulation with respect to scintillation effects.

There are two major limitations specific to these simulations, one of which this work seeks to mitigate. The multi-aperture approach cannot perfectly match the theoretical statistics for even a constant $C_n^2$ profile. This is due to an approximation by Taylor series at the center of its theoretical analysis \cite{Chimitt2020}. Within their analysis, it is \emph{impossible} to simulate with the multi-aperture simulation for a varying $C_n^2$ profile. These limitations in accuracy and the inability to simulate varying turbulent profiles are what we overcome in this work. Swapping their correlation statistics for ours within their simulation, both constant and path-varying turbulence profiles can be matched both in theory and empirically. The core benefit here is that multi-aperture simulation is orders of magnitude faster than split-step allowing for large amounts of data generation.

\subsubsection{Alternative Simulation Approaches}
There are many other approaches that seek to model the effects generated by the atmosphere on a wave or image. One of these alternatives is known as the brightness function simulation developed through a series of works \cite{Vorontsov_2005_a, Lachinova_2007_a, Lachinova_2017_a}. The brightness function model is faster than split-step, instead propagating ``bundles'' of rays through a perturbing medium. These bundle of rays are then distributed across the imaging plane for each pixel as a function of the medium, resulting in spatially varying effects as a function of the phase screens. More traditional ray tracing approaches beyond the brightness function model have been applied for the simulation and modeling of turbulent effects. Voelz et al. \cite{Voelz_2018_a} provides an analysis of standard ray tracing approaches, with carefully performed ray tracing matching wave optics simulations to a suitable degree of accuracy for most applications. Additionally, a comprehensive work on a similar simulation modality is described by \cite{Peterson_2015_a} and made publicly available.

In addition to these methods, there exist simulation approaches by Repasi and Weiss \cite{Repasi_2008_a, Repasi_2011_a}, Leonard et al. \cite{Leonard_2012_a}, or Potvin et al. \cite{Potvin_2011_a} which use a blend of analytic and empirical properties (empirically based on the NATO RTG-40 dataset \cite{Tofsted_2006_a, Tofsted_2007_a}) to simulate PSFs and the subsequent images directly. These simulation methodologies have been revisited more recently by Miller et al. \cite{Miller_2019_a, Miller_2021_a}. These methods share some similarity to Zernike-based methods, with a main difference being that the coefficients drawn in order to simulate the effects on an image describe various quantities in the spatial domain as opposed to the phase domain.

\subsection{Angle-of-Arrival Correlations}
\begin{figure}
    \centering
    \includegraphics[width=0.9\linewidth]{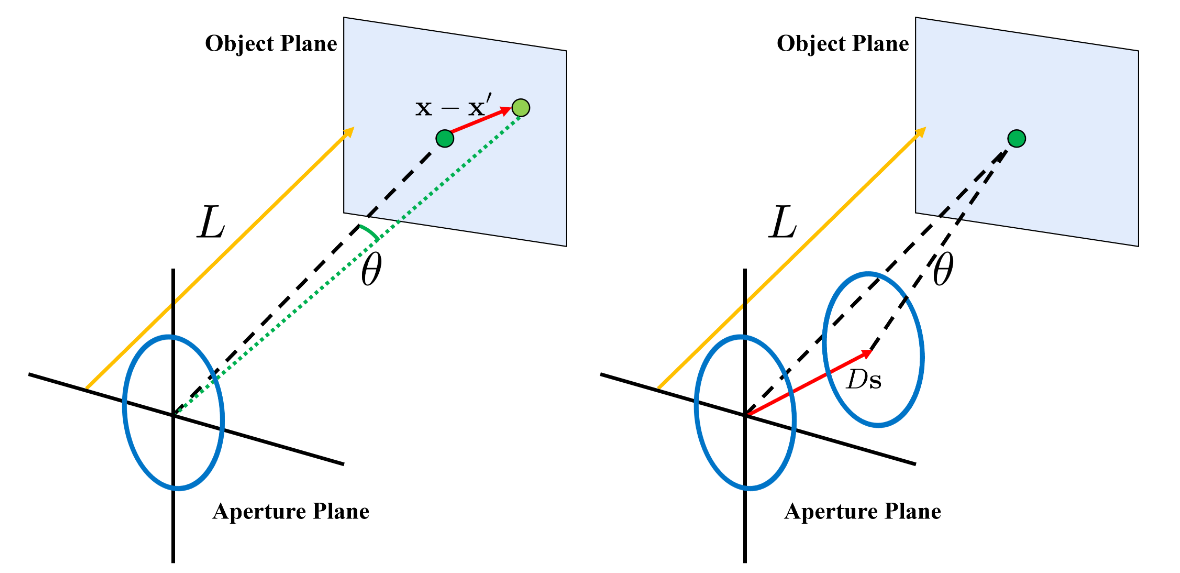}
    \begin{tabular}{ccc}
        (a) Angle-of-Arrival Geometry & \hspace{5ex} & (b) Two Aperture Geometry
    \end{tabular}
    \caption{Visualization of the geometries of the two types of problems. In (a) we show the problem analyzed by Fried \cite{Fried1976_variety}, where two points in the object plane located at positions $\vx, \vx'$. In (b) we show the problem of Takato and Yamaguchi \cite{Takato1995} where two apertures, with separation $D\vs$, are viewing a single point source in the object plane.}
    \label{fig: fried_vs_takato}
\end{figure}

The primary theoretical comparison used for verification of our approach is that of angle-of-arrival correlations as performed by Fried \cite{Fried1976_variety}. Tilt has a long correlation range, and is an important marker for accurate generation. Fried analyzed the correlation of distortions of two separate point sources in the object plane. As separation between the two points increases, we expect them to be less correlated as they are propagating through increasingly different regions of the atmosphere. In this context, Fried specifically analyzed the tilt vector, the vector normal to the plane of best fit for phase distortion $\phi_\vx(\vrho)$. The tilt vector is defined to be
\begin{equation}
    \valpha_{\vx} = \frac{2\lambda}{R} \int d\vrho P(\vrho) \phi_\vx(R\vrho) \vrho.
\end{equation}
Preferring to write this consistently with our usage of the Zernike polynomials, we may write
\begin{equation}
    \valpha_{\vx} = \frac{\lambda}{R} \int d\vrho P(\vrho) \phi_\vx(R\vrho) [Z_2(\vrho) \hat{\vi} + Z_3(\vrho) \hat{\vj}],
\end{equation}
with unit vectors $\hat{\vi}$ and $\hat{\vi}$ in the x and y directions, accordingly.
The problem of finding the tilt correlation can then be written compactly using our notation as
\begin{equation}
    \E [\valpha_{\vx}^T \valpha_{\vx'}] = \left(\frac{\lambda}{R}\right)^2 \left( \E [a_{\vx,2} a_{\vx',2}] + \E [a_{\vx,3} a_{\vx',3}] \right),
    \label{eq: tilt_corr_zernike_approx}
\end{equation}
where we note the $x$-tilt and $y$-tilt terms are independent, therefore there is no cross correlation to account for. We visualize this problem in \fref{fig: fried_vs_takato}(a). We note this may also be written using the phase structure function, however with the added consideration of the separation of the point sources,
\begin{align}
    \calD(R\vrho - R\vrho', \vx - \vx')& = 2.91 k^2 \int_0^L dz C_n^2(z)  \left| R(\vrho - \vrho') + \left(\frac{z}{L - z}\right)(\vx - \vx') \right|^{5/3}.
    \label{eq: struct_fun_sphere_two_points}
\end{align}
The magnitude term in this expression may be viewed as the varying distance between the two difference vectors as a function of $z$.

Fried's results are limited to the two tilt Zernike functions. For our application, if one desires to pull statistics for the higher order aberrations, the spatial correlation functions are required for generation. Additionally, this result only provides a description of the joint behavior of the Zernike coefficients. Therefore, additional work will be required to empower multi-aperture methods.

\subsection{Two-Aperture Correlations}
We now turn to a slightly different problem, though as we will show, closely related. The works of Chanan \cite{chanan92} and more generally Takato and Yamaguchi \cite{Takato1995} analyze the correlation of two \emph{different} imaging systems both imaging the same point. These are inherently different from angle-of-arrival as they are imaging the same point source. These works were inspired by astronomical imaging situations in which two apertures can give additional useful information on the correlation of the wavefront distortions allowing for improved distortion correction.

Takato and Yamaguchi as well as Chanan consider apertures separated by a vector $\vs$ measured from center to center and is written in terms of the aperture size $D$. For example, $D\vs$ with $\vs=1$ corresponds to two apertures that are tangential. In our notation, their problem can be written as
\begin{align}
    \E [a_{\vx,i} (\vec{0}) a_{\vx,j} (\vs)] = \frac{1}{\pi^2} \iint d\vrho d\vrho' P(\rho) P(\rho') Z_i(\vrho) Z_j(\vrho') \E[\phi_{\vx}(R\vrho) \phi_{\vx}(R\vrho' + D\vs)].
    \label{eq takatos_problem}
\end{align}
We provide a visualization of this problem in \fref{fig: fried_vs_takato}(b). We note that this problem is equivalent to Noll's in the case of $\vs = \Vec{0}$. This formulation may be used in accordance with the structure function \eqref{eq: struct_fun_sphere} with additional displacement $D\vs$. The resulting expression from the analysis of \cite{Takato1995} is rather cumbersome, therefore, we take some care to simplify the notation with the intent of clarity which we leave to Appendix \ref{ap: takato}. Their final result is given by
\begin{align}
    \E [a_{\vx,i} (\vec{0})  a_{\vx,j} (\vs)] = 0.00969  k^2 2^{14/3} \pi^{8/3} (D/2)^{5/3} \sqrt{(n_i + 1)(n_j + 1)} f_{ij} (\vs, k_0) \int dz C_n^2(z),
    \label{eq: takato_main_result}
\end{align}
giving the correlation for two apertures separated by a vector $\vs$, with $(n_i,m_i) \to i$, and similarly for $j$. We note that this may be written in terms of $D/r_0$ with $r_0$ as the Fried parameter\cite{Fried66optical, roggemann1996imaging}, defined as
\begin{equation}
r_0 = 0.185\left[\frac{4\pi^2}{k^2 \int_0^L \left(\frac{L - z}{L}\right) C_n^2(z)}\right]^{3/5},
\end{equation}
where we have chosen to write the spherical form (the planar form drops the $(L - z)/L$ term in the integral). 
The expression describing $f_{ij}$ is given in the Appendix \ref{ap: takato}. 

\section{Deriving the Spatial Zernike Correlations}
The two groups of Zernike spatial correlations considered, angle-of-arrival and two-aperture correlations, are for two separate groups of problems. However, to enable the generation of the optical statistics by methods of \cite{Chimitt2020, Mao_2021_ICCV}, we merge the two ideas to describe Zernike correlations for all coefficients. We begin with stating the problem at hand as
\begin{align}
    \E [a_{\vx,i} a_{\vx',j}] = \frac{1}{\pi^2} \iint d\vrho d\vrho' P(\rho) P(\rho')  Z_i(\vrho) Z_j(\vrho') \E[\phi_{\vx}(R\vrho) \phi_{\vx'}(R\vrho')].
    \label{eq: our_problem}
\end{align}
This differs from \eqref{eq takatos_problem} by consideration of two point sources located at points $\vx, \vx'$.
As before, this may be written using the phase structure function as
\begin{align}
    \E [a_{\vx,i} a_{\vx',j}] = \frac{-1}{2\pi^2} \iint d\vrho d\vrho'  P(\rho) P(\rho')  Z_i(\vrho) Z_j(\vrho')  \calD(R\vrho - R\vrho', \vx - \vx').
    \label{eq: our_problem_with_struct}
\end{align}
The formulation of our problem is then most in accordance with Fried's approach, though notably we have changed from the case of the tilt vector to any arbitrary Zernike polynomial.

This section begins with presenting the main theoretical result of this work: the correlation of the  Zernike coefficients for the case of a continuous turbulence profile. This will consist of the usage of \eqref{eq: takato_main_result} to solve \eqref{eq: our_problem}. We will then move to discretize the main continuous results, causing us to define a $C_n^2$-slice, which will draw some analogy to a phase screen from split-step. This second perspective will be the one taken for the sake of numerical evaluation. We finish with how the approximation of Chimitt and Chan \cite{Chimitt2020} fits into this framework and discuss its limitations.

\subsection{Continuous Case: Varying $C_n^2$}
Before carrying out the main derivation, we must mention that our approach will be applied to the case of spherical waves. However, the results of Takato and Yamaguchi are developed for planar waves. This creates a potential conflict, applying results from planar waves to spherical ones. To remedy this, we facilitate the general nature of their results which has no restriction on the $C_n^2$ profile. As shown in \eqref{eq: takato_main_result}, the turbulence profile is a non-closed expression in their final result. We may therefore choose to write the turbulence profile to satisfy our requirements via
\begin{equation}
    C_n^2(z) =  \left(\frac{L - z}{L}\right)^{5/3}
    \tilde{C}_n^2 (z) [ u(z) - u(z - L) ],
\end{equation}
where $u(z - a)$ is the unit step function which is unity for $z > a$ and 0 otherwise and $\tilde{C}_n^2 (z)$ is the original, unmodified turbulence profile in the planar case (of course, the two are the same). Substitution of this turbulence profile into \eqref{eq: takato_main_result} then satisfies our requirement for spherical waves. Our later comparisons will be done for spherical wave statistics, which lends credence to this approach.

With our approach to using Takato and Yamaguchi's results for spherical waves in mind, we first begin with rewriting \eqref{eq: struct_fun_sphere_two_points} as
\begin{align}
    \calD(R\vrho - R\vrho', & \vx - \vx') = 2.91 k^2 \int_0^L dz\left(\frac{L - z}{L}\right)^{5/3} C_n^2(z) \left| R(\vrho - \vrho') + \left(\frac{z}{L - z}\right)(\vx - \vx') \right|^{5/3}.
\end{align}
This allows us to write \eqref{eq: our_problem_with_struct} as
\begin{align}
    \E & [a_{\vx,i} a_{\vx',j}] = \frac{-2.91 k^2}{2\pi^2} \int dz\left(\frac{L - z}{L}\right)^{5/3}  C_n^2(z) \iint d\vrho d\vrho' \notag\nonumber\\
    \times & P(\rho) P(\rho') Z_i(\vrho) Z_j(\vrho') \left|R(\vrho - \vrho') + \left(\frac{z}{L - z}\right)(\vx - \vx')\right|^{5/3}.
    \label{eq: our_problem_pre_result}
\end{align}
If we seek to leverage the results of the two-aperture statistics, we must relate the magnitude term to some two-aperture separation in accordance with Takato and Yamaguchi. We therefore define
\begin{equation}
    \vs(z) = \left(\frac{z}{D(L - z)}\right)(\vx - \vx'),
    \label{eq: s_displacement}
\end{equation}
to be a displacement that is changing with distance along the path of propagation. With this substitution, we can write
\begin{align}
    \E [a_{\vx,i} & a_{\vx',j}] = \frac{-2.91 k^2}{2\pi^2} \int dz\left(\frac{L - z}{L}\right)^{5/3}  C_n^2(z) \iint d\vrho d\vrho'   \nonumber\\
    \times & P(\rho) P(\rho') Z_i(\vrho) Z_j(\vrho') \left|R(\vrho - \vrho') + D\vs(z)\right|^{5/3}.
\end{align}

\begin{figure}
    \centering
    \includegraphics[width=0.9\linewidth]{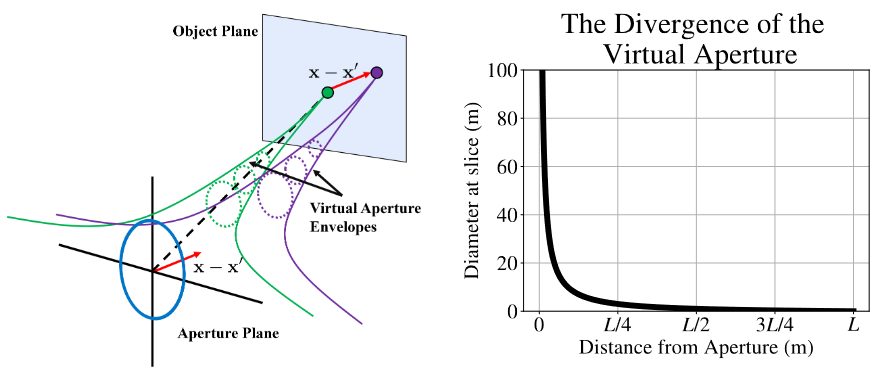}
    \begin{tabular}{ccc}
        (a) Cone & \hspace{30ex} & (b) Side
    \end{tabular}
    \caption{Visualization of the virtual aperture. In (a) we show in 3D space how the virtual aperture changes with position along the length of propagation. In (b) we show the increase in the virtual aperture diameter as a function of position.}
    \label{fig: virtual_aperture_figure}
\end{figure}

We then recognize the inner double integral to be of the same form as \cite{chanan92, Takato1995}. Defining $\mathscr{A} = 0.00969 k^2 2^{14/3} \pi^{2/3} R^{5/3}$, this allows us to simply leverage the results of Takato and Yamaguchi using a weighted integration of their solutions, resulting in
\begin{align}
    \E[a_{\vx,i} a_{\vx',j}] =  \mathscr{A}_{i,j}
    \int_0^L \left(\frac{L - z}{L}\right)^{5/3} C_n^2(z) f_{ij} \left( \vs(z), k_0 \right) dz,
    \label{eq: main_result_continuous}
\end{align}
where $\mathscr{A}_{i,j} = \mathscr{A} \sqrt{(n_i + 1)(n_j + 1)}$.
This result, however, has an additional visual interpretation. Turning to \eqref{eq: s_displacement}, we can write this in terms of a ``virtual'' aperture which varies with distance, which we define to be
\begin{equation}
    \hat{D}(z) = D \left(\frac{L - z}{z}\right).
\end{equation}
We may then write the displacement as $\vs(z) = (\vx - \vx')/\hat{D}(z)$. We provide a visualization of this virtual aperture in \fref{fig: virtual_aperture_figure} which illustrates two points in an object forming two cones with diverging radii. The overlap of the cross sections at each individual infinitesimal slice is the problem analyzed by Takato and Yamaguchi. However, our result differs by being a \emph{sum} of these solutions; each infinitesimal slice contributes a correlation which is dictated by the results of \cite{Takato1995}. Intuitively, the slice closest against the aperture will contribute global correlation. Physically this is due to the fact that every point source on an object will pass through this slice. Mathematically, the case of this final slice will result in $\hat{D}(z) \to \infty$, which will cause $\vs(z) \to 0$ for all finite $\vx - \vx$, implying perfect global correlation. To present the main result completely, we substitute \eqref{eq: s_displacement} into \eqref{eq: main_result_continuous}, giving us
\begin{align}
    \E[a_{\vx,i} a_{\vx',j}] =  \mathscr{A}_{i,j}
    \int_0^L \left(\frac{L - z}{L}\right)^{5/3} C_n^2(z) f_{ij} \left( \left(\frac{z}{D(L - z)}\right)(\vx - \vx'), k_0 \right) dz.
    \label{eq: main_result_continuous2}
\end{align}

\subsection{Discrete Case: $C_n^2$-slices}
\label{sec: cn2_slice}
A unique perspective given in this work is the direct application to simulation of turbulent optics statistics. The expression for the correlation of two points in the object plane \eqref{eq: main_result_continuous2} is desirable for the sake of simulation. These results can be directly used in previous multi-aperture simulations \cite{Chimitt2020, Mao_2021_ICCV} with only minor modification. In addition to this, \eqref{eq: main_result_continuous2} is preferable to the results of Whiteley et al. \cite{Whiteley98} for numerical evaluation of the integral. This is due to the fact that their derivation used a particular Riemann summation rule as a component of their derivation. Equation \eqref{eq: main_result_continuous2} assumes no such rule. The result is an expression which can be adapted to suit the needs of the application without needing to re-derive for a separate integration rule. In particular, the way in which one performs a summation over $C_n^2(z)$ is the primary knob one may tune to utilize the results of \eqref{eq: main_result_continuous2}.

Since one may choose the way in which the integral is represented as a Riemann sum, one possibility would be to average the turbulence profile along the interval of propagation. There is some analogy here to the phase screens used in split-step propagation, though they are not directly equivalent. The main difference is that within this framework, one does not have to actually generate phase screens, rather, just evaluate the statistical expression. To differentiate the two, we denote them as $C_n^2$-slices.
We begin with defining a collection of $C_n^2$-slices to be
\begin{equation}
    C_n^2(z; M) = \sum_{m=1}^M \delta\left(z - \frac{Lm}{M+1}\right) \int_{L(m-1)/M}^{Lm/M} C_n^2 (v) dv,
    \label{eq: cn2_slices_sum}
\end{equation}
where we are integrating along the path of propagation using $v$ as a dummy variable. For some simplicity in notation, we define the locally collapsed $m$th $C_n^2$-slice to be
\begin{equation}
    \overline{C_n^2}(z_m) = \int_{L(m-1)/M}^{Lm/M} C_n^2(v) dv.
\end{equation}


Using the representation of $C_n^2$-slices in place of the turbulence profile in \eqref{eq: main_result_continuous2}, we can arrive at
\begin{align}
    \E[a_{\vx,i} a_{\vx',j}; M] = \mathscr{A}_{i,j} & \sum_{m=1}^M \left(\frac{M+1-m}{M+1}\right)^{5/3} \overline{C_n^2}(z_m) f_{ij}\left(\frac{m (\vx - \vx')}{D(M+1-m)}, k_0\right).
    \label{eq: main_result_discrete}
\end{align}
We use \eqref{eq: main_result_discrete} as a basis for our numerical testing. Due to the generality of the expression provided in \eqref{eq: main_result_continuous2}, one may use an alternative integration rule in order to decide each $C_n^2$ value as previously stated. Furthermore, instead of focusing on numerical integration, one may instead optimize the discrete $C_n^2$ values for objective functions which optimize quantities such as the Fried parameter, isoplanatic angle, and log amplitude variance as in Hardie et al. \cite{HardieSimulator}. In this case, individual $C_n^2$ values which describe the phase screen parameters were optimized in a fashion similar to that by Schmidt \cite{SchmidtTurbBook}. Thus \eqref{eq: main_result_discrete} can also be used to evaluate the impact of simulation parameters on the various Zernike correlations if one replaces $\overline{C_n^2}$ with other values.

\subsection{Comparison to Earlier Multi-Aperture Methods}
The fundamental work that enabled the first iteration of the multi-aperture simulation \cite{Chimitt2020} imposed two main restrictions to achieve their results. The first is the assumption of a constant $C_n^2$ profile. However, the more limiting restriction is an approximation upon the structure function, which they use a first order Taylor series to simplify. The following consideration achieves the same results, shedding some light on the limitations of these previous results.

To arrive at these same results, we first choose to define a single $C_n^2$-slice along the entire path of propagation,
\begin{equation}
    C_n^2(z; 1) = \int_0^L \delta\left(z - \frac{L}{2}\right) C_n^2 (v) dv.
\end{equation}
With the additional assumption of a constant $C_n^2$ profile, then
\begin{equation}
    C_n^2(z; 1) = L C_n^2 \delta\left(z - \frac{L}{2}\right).
\end{equation}
The resulting substitution of this $C_n^2$ profile results in the same correlation function as in \cite{Chimitt2020},
\begin{equation}
    \E[a_{\vx,i} a_{\vx',j}; 1] = \mathscr{A}_{i,j}  \left(\frac{1}{2}\right)^{5/3} L C_n^2 f_{ij}\left(\frac{ (\vx - \vx')}{D}, k_0\right).
    \label{eq: chim_chan_main_result_1}
\end{equation}
This demonstrates the limiting assumption inherent to previous multi-aperture simulators more clearly. Specifically, these simulation methodologies assume a single $C_n^2$-slice at the halfway point of the propagation path. Beyond previously discussed limitations, even the case of constant $C_n^2$ profiles will experience moderate deviations given certain camera configurations, which has been observed \cite{Chimitt2020}. To comment more directly on the usage of our results within theirs, the more general results in this work only requires the replacement of \eqref{eq: chim_chan_main_result_1} with \eqref{eq: main_result_discrete}, keeping the rest of the simulation the same. This offers a considerable increase in accuracy with only an initial small loss in speed -- after the generation of the theoretical spatial correlation, the generation method (and therefore the speed) is identical.

\subsection{Comparison to Existing Zernike Correlations}
The work most closely aligned with ours is that of Whiteley et al. \cite{Whiteley98}. Whiteley et al. provides an expression for spatial and temporal Zernike correlations, and similarly utilize Takato and Yamaguchi. The core difference is notably in approach taken and result. On the approach side, \cite{Whiteley98} uses an assumption on the independence of neighboring atmospheric slices. Ours additionally uses such a fundamental assumption, as this underlies the standard Markov approximation, however our inclusion of this independence is ``built into'' the result of Takato et al. \cite{Takato1995}. Therefore, it is useful to make the comment that both works ultimately rely on this same layered atmospheric property -- though Whiteley et al. \cite{Whiteley98} assumes this earlier in their derivation. The result is that if one wishes to change a Riemann sum approach (such as max/min vs. Simpson's rule) one must carry the derivation out again. With ours, one may start at the result of \eqref{eq: main_result_continuous2}, and discretize as desired following Section \ref{sec: cn2_slice} directly. Furthermore, the visual interpretation of the integration process via the virtual aperture along with the subjectively more convenient form of \eqref{eq: main_result_continuous2} are two added benefits of our approach.

It is important to note that Whiteley et al. \cite{Whiteley98} does offer a more general framework with regards to aperture motion and temporal correlations. This is done primarily through use of Taylor's frozen flow hypothesis. To illustrate how these concepts may be incorporated into our framework, the temporal effects can be modeled in a similar fashion via
\begin{equation}
    \vs(z) = \left(\frac{z}{D(L - z)}\right)(\vx - \vx') + \frac{\vv(z) \tau}{D},
    \label{eq: s_t_displacement}
\end{equation}
where $\vv(z)$ is the mean transverse wind velocity at position $z$ along the path of integration. We note the division by $D$ is a requirement to match the form of Takato and Yamaguchi's expression \cite{Takato1995}. We may substitute \eqref{eq: s_t_displacement} into \eqref{eq: main_result_continuous}, along with potential modifications of \eqref{eq: s_t_displacement} as outlined in Sasiela's book \cite{sasiela2007electromagnetic}.

\section{Numerical Comparisons and Discussions}
Our numerical results can be divided into two sections: theoretical comparisons and empirical statistics. On the side of theoretical comparisons, we first numerically compare the angle-of-arrival results with our expression. We present a wide variety of situations, all of which we are able to match sufficiently close. We accept this as a strong suggestion of their equivalence, though the generality of the expression obtained in this work makes a direct analytical equivalence difficult. Furthermore, we discuss some difficulties that arise in split-step when simulating slant paths and other complicated $C_n^2$ profiles and how this is described within our model. All of the following computations were performed on a AMD Ryzen 5 3600 6-core CPU with 16 GB or RAM on a 64-bit OS. Our implementation specifically utilizes Python and the numpy/scipy libraries.

On the empirical side, we compare the speeds of split-step and the multi-aperture method for simulation of a grid of point sources. We then demonstrate the limitation in resolution for Zernike-based simulations and its reason, as well as outlining general principles towards a solution, with more details in Chimitt et al.\cite{Chimitt_2022}.

\subsection{Comparisons to Angle-of-Arrival Results}
\label{sec: compare_ma_vs_aoa}
To begin our comparisons, we first look at the comparison between previously reported angle-of-arrival results \cite{Basu_2015, Fried1976_variety} with our expression \eqref{eq: main_result_discrete}. Recalling \eqref{eq: tilt_corr_zernike_approx}, we note that the tilt expression in our framework is proportional to the sum of the Zernike tilt coefficient variances. Therefore, we may write the result for angle-of-arrival using our results by summation of the terms corresponding to the tilt Zernike terms within \eqref{eq: main_result_discrete}. We contrast the analytical form of the correlation here to those given in \cite{Basu_2015, Fried1976_variety}. We observe this to be a nearly identical match between the expressions provided in these two works. These expressions are developed for the case of spherical waves, matching our development.

To compare the two, we can evaluate the angle-of-arrival integral directly using the Python library scipy's integration class. Specifically, we use the triple integration method `tplquad', writing the angle-of-arrival expression with lambda functions. Therefore, no elements of our $C_n^2$-slice concept are a component of the angle-of-arrival integral evaluation. For generating the curves predicted by our analysis, we may instead evaluate \eqref{eq: main_result_discrete}. To save on a bit of time, we find the values for the $C_n^2$-slices by taking $10\times$ the number of points needed and averaging over groups of ten (thus an approximation of \eqref{eq: cn2_slices_sum}). For example, if we require 10 $C_n^2$-slices, these values are estimated from 100 samples of the $C_n^2$ profile via local averaging.

\begin{figure}
    \centering
        \includegraphics[width=0.75\linewidth]{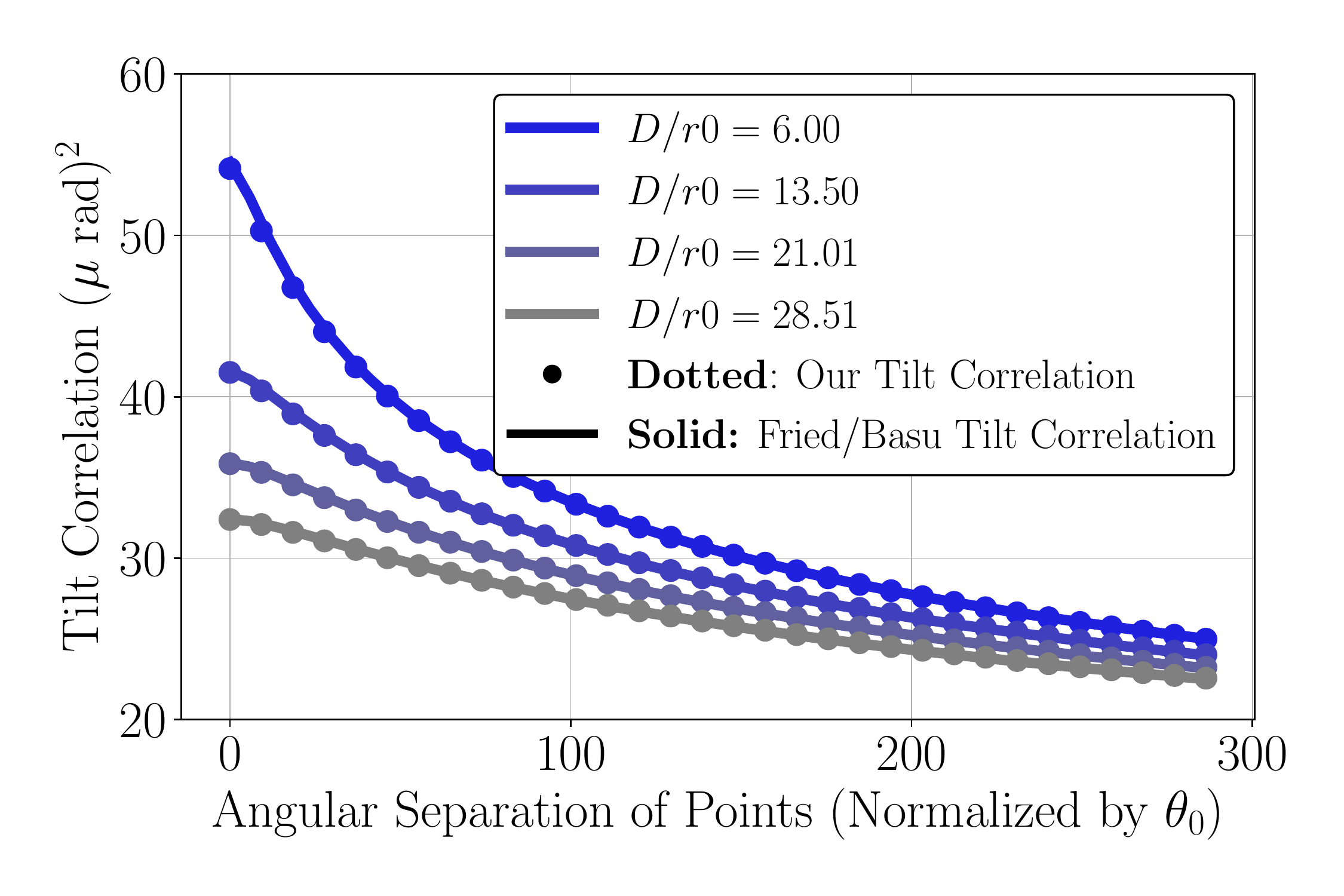}
    \caption{Theoretical comparison of our expression for tilt correlation compared to Basu et al. (now Bose-Pillai) \cite{Basu_2015} for varying aperture diameters. These results are for a constant $C_n^2$ profile of $C_n^2 = 2 \times 10^{-15} \text{ m}^{-2/3}$.}
    \label{fig: constant_cn2}
\end{figure}

\begin{figure}
    \centering
        \includegraphics[width=0.75\linewidth]{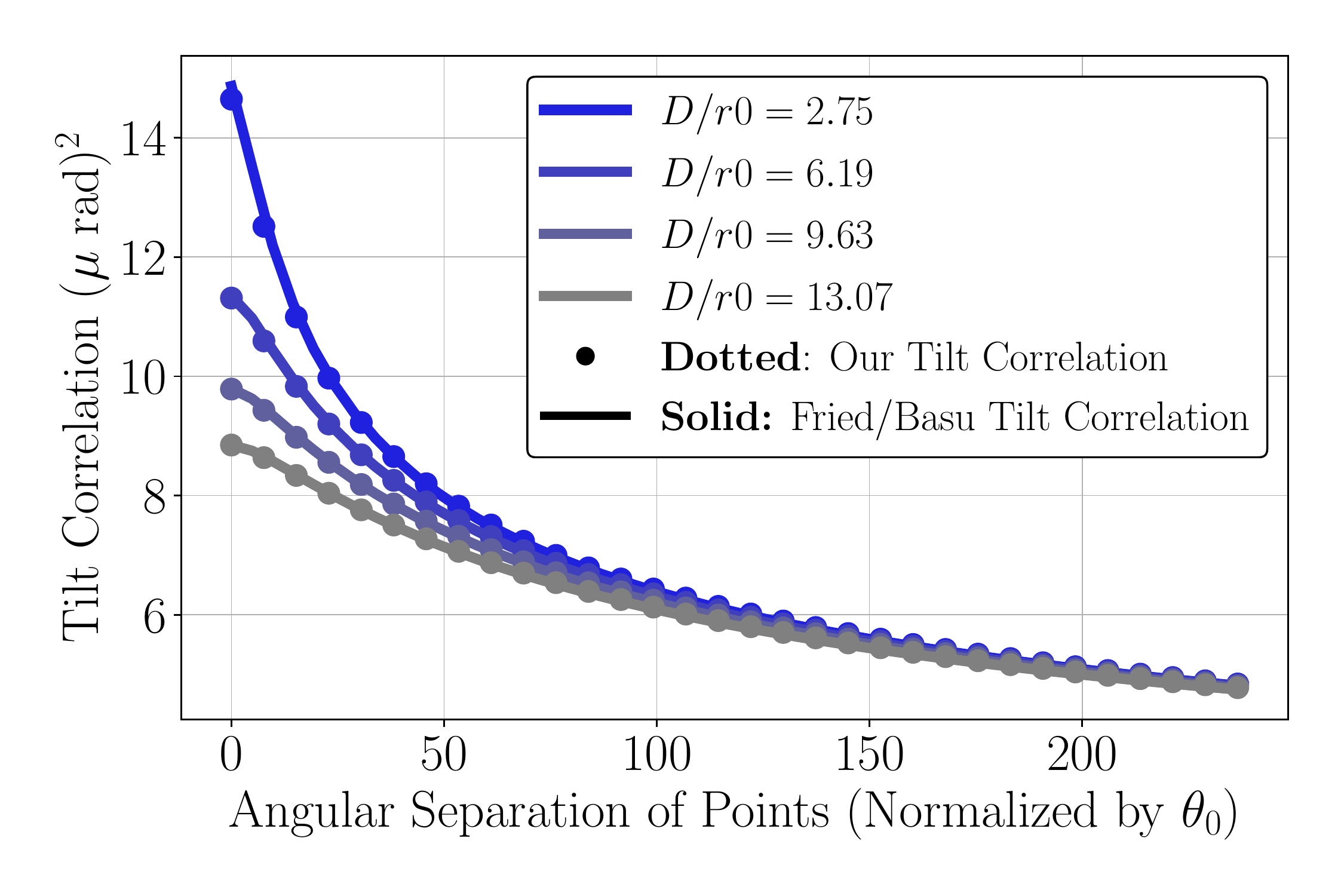}
    \caption{Theoretical comparison of the expression for our tilt correlation compared to \cite{Basu_2015}. The results shown is for a path-varying turbulence profile that is high at the aperture $C_n^2(z) = 2(z/L) \times 10^{-15} \text{ m}^{-2/3}$ and with varying aperture sizes.}
    \label{fig: varying_cn2}
\end{figure}

The results of the constant $C_n^2$ profile are presented in \fref{fig: constant_cn2}, while the path-varying turbulence profile results can be seen in \fref{fig: varying_cn2}. Here we normalize by the isoplanatic angle on the x-axis, given as
\begin{equation}
\theta_0  = 58.1 \times 10^{-3} \lambda^{6/5} \left[\int_0^L z^{5/3} C_n^2(z) dz\right]^{-3/5}.
\end{equation}
We observe a convincing match for all parameters within our tests. That is, the angle-of-arrival integral appears to be similar to the results predicted by this analysis. We further note that all evaluations of \eqref{eq: main_result_discrete} were performed with the number of $C_n^2$-slices kept constant at 200. We have noticed some minor improvement with an increase of slices (i.e. our curves match the angle-of-arrival curves more closely), but we find 200 to be sufficient for the purposes of this comparison. We again note that the angle-of-arrival results from Ref. \cite{Basu_2015} come from a separate analysis and are evaluated with methods separate from ours.

\subsection{The Problem with Small Numbers of Phase Screens}
When analyzing or simulating propagation through turbulence using the discrete phase screen approach \cite{roggemann1996imaging}, the number of phase screens can be chosen to match various requirements of the application. For analysis which utilizes phase screens, the number of phase screens is typically left to be unspecified so long as assumption of phase screen independence holds. In simulation, 10 phase screens are used in Hardie et al. \cite{HardieSimulator} which we deem to be relatively standard, though one may use more/less if the situation dictates. This leads to a difference in analysis vs. simulation: one uses a large number of phase screens whereas in simulation a small number is used.

To quantify this difference in analysis and simulation, one aspect we may study is the tilt correlation. We may ask: Does a simulation with a small amount of phase screens match the analytic prediction with a larger amount of independent phase screens? To answer this question without reliance on empiricism, we first note that a properly performed simulation should perform identically to its analytic counterpart. Therefore, we may analyze simulation by a representative analytical model.

The model we choose for our purposes is similar to the $C_n^2$-slice model, though the values of the Riemann sum terms are chosen to optimize the objective function provided in Hardie et al. \cite{HardieSimulator}. We note that this was done to choose phase screen parameters to closely match isoplanatic angle, Fried parameter, and log amplitude variance in a least-squares fashion. Thus, the model we choose is
\begin{equation}
    C_n^2(z) = \sum_{m=1}^M \delta\left(z - \frac{Lm}{M+1}\right) \Tilde{C}_{n,m}^2,
    \label{eq: phase_screen_model}
\end{equation}
where $\Tilde{C}_{n,m}^2$ is the optimally chosen $C_n^2$ value of the $m$th phase screen (optimized according to Hardie et al. \cite{HardieSimulator}). We will refer to \eqref{eq: phase_screen_model} as the phase screen model.

\begin{figure*}
    \centering
    \includegraphics[width=0.9\linewidth]{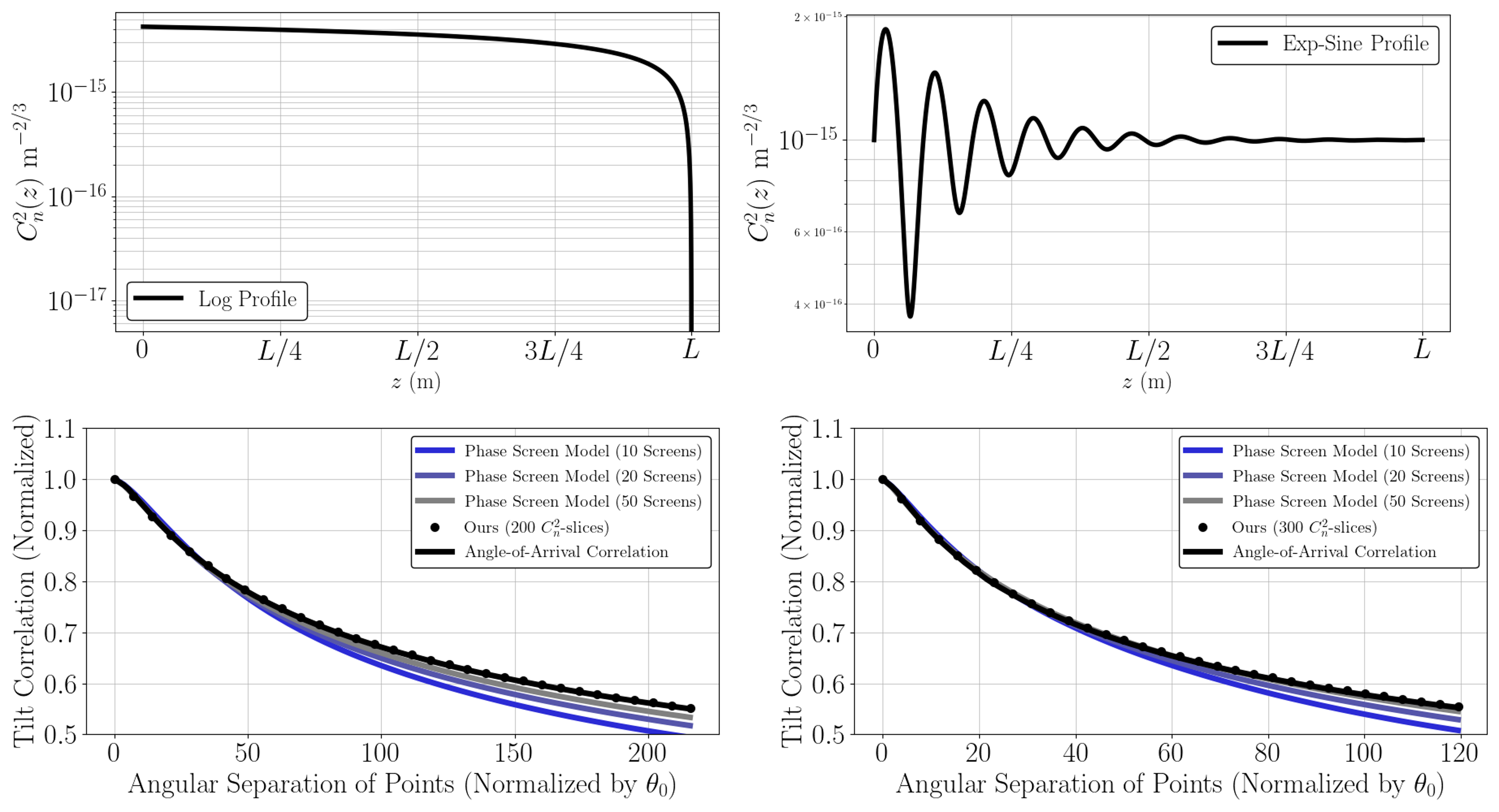}
    \begin{tabular}{ccc}
        (a) $C_n^2(z) = 1\times 10^{-15}$ & \hspace{5ex} & (b) $C_n^2(z) = 1 \times 10^{-15}$ \\
        $\times \log((z-L)/100 + 1) \text{ m}^{-2/3}$ & \hspace{5ex} & $\times(e^{-z/1000} \sin(z/100) + 1) \text{ m}^{-2/3}$  \\
    \end{tabular}
    \caption{[Top] For two different $C_n^2$ profiles (shown in the log domain) we can plot the [Bottom] tilt correlation for each using (1) angle-of-arrival correlations, (2) $C_n^2$-slice correlations, (3) the phase screen model correlations. We note that the phase screen model's $C_n^2$ values are optimized as described in Hardie et al.\cite{HardieSimulator}. We can see that with increasing the number of phase screens, the phase screen model approaches the angle-of-arrival and high $C_n^2$-slice curves.}
    \label{fig: ma_hard_cn2_1}
\end{figure*}

\begin{figure*}
    \centering
    \includegraphics[width=0.9\linewidth]{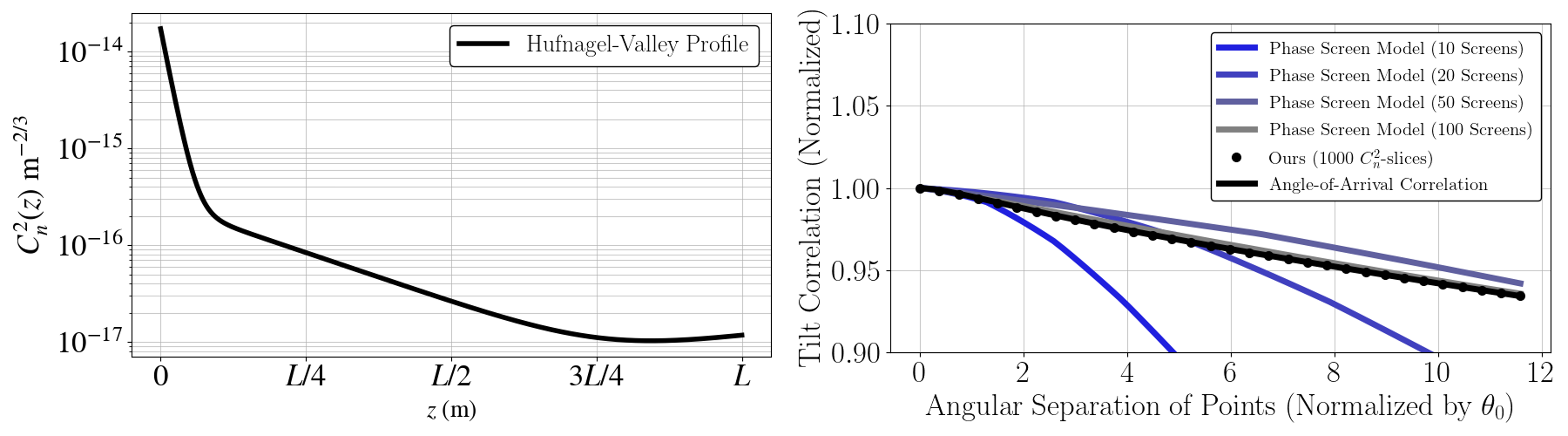}
    \begin{tabular}{ccc}
        (a) Hufnagel-Valley Profile & \hspace{10ex} & (b) Empirical Correlations
    \end{tabular}
    \caption{[Left] For the Hufnagel-Valley $C_n^2$ profiles (shown in the log domain) we can plot the [Right] tilt correlation for each using (1) angle-of-arrival correlations, (2) $C_n^2$-slice correlations, (3) the phase screen model correlations. We note that the phase screen model's $C_n^2$ values are optimized as described in Hardie et al.\cite{HardieSimulator}. We can see that with increasing the number of phase screens, the phase screen model approaches the angle-of-arrival and high $C_n^2$-slice curves.}
    \label{fig: ma_hard_cn2_2}
\end{figure*}

Therefore, studying the impact of $M$ on the accuracy of the tilt correlation may be studied analytically as we have the tilt correlation integral \eqref{eq: main_result_continuous2} and the phase screen model \eqref{eq: phase_screen_model} for this case. We present the results for two somewhat challenging cases in \fref{fig: ma_hard_cn2_1} and \fref{fig: ma_hard_cn2_2}. In these cases, we provide both the angle-of-arrival correlations\cite{Basu_2015} (via direct scipy integration) for reference along with the case of 200 $C_n^2$-slices using \eqref{eq: main_result_discrete}. In this case, we observe that if we are interested in a narrow field of propagation (corresponding to a small value on the x-axis of \fref{fig: ma_hard_cn2_1} and \fref{fig: ma_hard_cn2_2}) then we may use a small number of phase screens to model the situation. However, if we are concerned with proper correlations in the more anisoplanatic case, a larger number of phase screens may be required depending on the desired accuracy.

In addition to these more complicated cases, we also use simpler profiles which match those of Hardie et al. \cite{HardieSimulator}. We find there to be a match with their reported empirical results when using their reported values for $\Tilde{C}_{n,m}^2$ within the phase screen model \eqref{eq: phase_screen_model}. We show these examples for two reasons: (1) we use this as supporting evidence that \eqref{eq: phase_screen_model} correctly models a properly performed split-step simulation; (2) that errors in tilt correlation are not a fault of split-step, but rather a function of either the optimization chosen to select the $C_n^2$ parameters or number of phase screens. With respect to (2), we note this particularly as stated in Hardie et al. \cite{HardieSimulator} that the mismatch in their reported tilt correlation was to be investigated in future work. Along similar lines as previously, the number of phase screens may be increased in order to match the tilt correlation function more closely.

\begin{figure}
    \centering
        \includegraphics[width=0.9\linewidth]{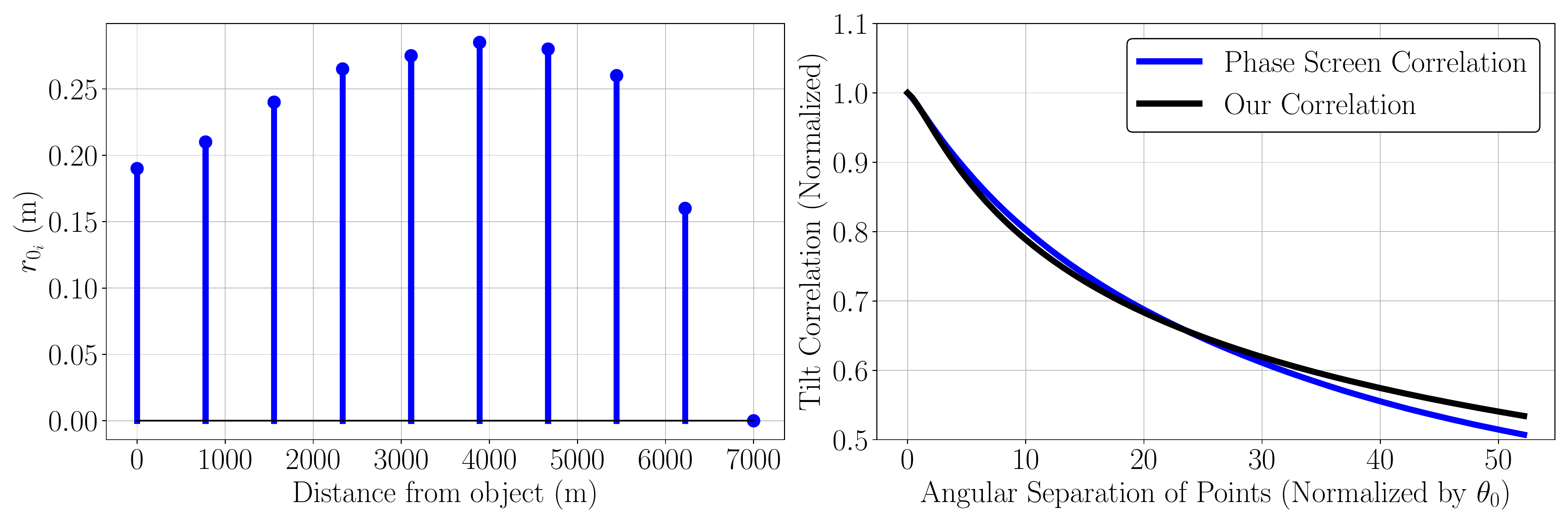}
    \caption{[Left] A set of phase screen $r_0$ values which model the $C_n^2$ profile, $C_n^2 = 0.25\times 10^{-15} \text{ m}^{-2/3}$ as given by Hardie et al.\cite{HardieSimulator}. [Right] The deviation in evaluation of our correlation integral with the phase screen model matches previously reported results\cite{HardieSimulator}.}
    \label{fig: ma_vs_ss_constant}
\end{figure}

\begin{figure}
    \centering
        \includegraphics[width=0.9\linewidth]{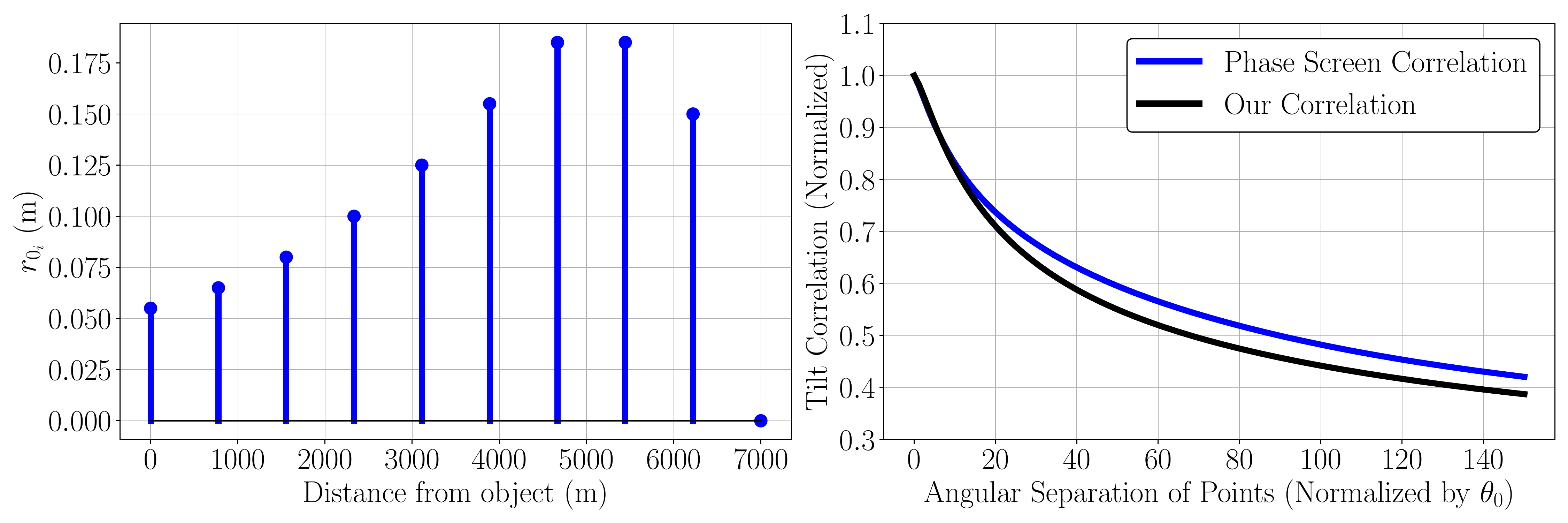}
    \caption{[Left] A set of phase screen $r_0$ values which model the $C_n^2$ profile, $C_n^2(z) = 2 (1 - z/L)$ as given by Hardie et al.\cite{HardieSimulator}. [Right] The deviation in evaluation of our correlation integral with the phase screen model matches previously reported results\cite{HardieSimulator}.}
    \label{fig: ma_vs_ss_varying}
\end{figure}

\subsection{Zernike-based Simulation with $C_n^2$-slices}

\subsubsection{Limitations and Approximations for Zernike-based Simulation}
To begin, we feel it important to discuss the limitations which one should keep in mind when interpreting the Zernike-based results. This variety of simulation is mainly motivated by speed -- the ability to sample points using FFT-based random sampling techniques \cite{Chimitt2020, Mao_2021_ICCV, Chimitt_2022} allows for a high degree of speed. The Zernike coefficient realizations, however, cannot directly be sampled with such FFT-based methods. The primary reason for this is that FFT-based sampling techniques require the assumption of wide-sense stationarity (WSS). It may seem that the previous results (i.e. Equation \eqref{eq: main_result_discrete}) may indeed be WSS. The WSS property requires that the correlation be a function of a difference, which is satisfied for $i = j$. To see this, one may note that \eqref{eq: main_result_discrete} is a function of spatial separation $\vx - \vx'$. However, when $i \neq j$, \eqref{eq: main_result_discrete} cannot be said to be WSS.

This imposes a serious restriction on Zernike-based simulation. We need \emph{multiple} coefficients to represent the turbulent phase distortions, of which 36 is the number chosen in some methods\cite{Mao_2021_ICCV, Chimitt_2022}. If we are only interested in simulating a handful of points (say, a $4 \times 4$ spatial grid) then this is not be too serious of a restriction, as Cholesky decomposition is possible in this case. Cholesky decomposition only requires the covariance matrix to be positive semi-definite, thus there is no issue with a lack of WSS. However, for a grid of spatial points in the object plane of size $H \times W$, one will require a correlation matrix the size of $ 36 HW \times 36 HW$ if using 36 Zernike coefficients\cite{Chimitt_2022}. This will often cause us encounter memory issues, a point we shall return to later.

At present, the way to circumvent this limitation is to utilize the fact that the covariance structure is near-diagonal \cite{Noll_1976}. If we may settle for an approximation, FFT-based generation may be allowed. This is a problem tackled in Chimitt et al. \cite{Chimitt_2022} in which the approximation takes on the form of a 2-stage FFT-based generation: independent coefficient generation followed by a mixing step. This approach is inspired by the fact that for a single coefficient the random process is WSS. The result is an approximation of the complete correlation structure of the Zernike. We wish to  highlight this for the following reason: in the speed comparisons we shall make, this approximation will be utilized. The loss in accuracy is quantified in Ref. \cite{Chimitt_2022}, with the speed gaining a factor of nearly $1000\times$ given certain configurations. Depending on the goal of the simulation, this may be an acceptable trade-off for a minimal drop in accuracy.

Finally, we wish to again remind the reader that aside from these sampling issues, the issue of generalization to amplitude effects is unclear. Therefore, split-step should still be regarded as more general with respect to the types of effects that may be incorporated into the methodology.

\subsubsection{Accuracy}
The accuracy of Zernike-based simulation depends on the (1) validity of the approximation of Chimitt et al.\cite{Chimitt_2022} and (2) whether or not the correlation kernels themselves are positive definite. The curves shown in previous Figures (such as \fref{fig: constant_cn2}, \fref{fig: varying_cn2}) will be empirically replicated upon averaging over random draws. The reason for this is that FFT-based generation is utilized -- as long as the functions are positive semi-definite, the empirical curves will match their analytic counterparts. We suggest the interested reader to Chimitt et al. \cite{Chimitt_2022} for more details regarding the approximation and its impact on the accuracy. That being said, for higher order Zernike correlations, one may expect slight deviation from the predicted curves due to the approximation (assuming one is using the described approximation).

An added benefit of the fact that previous versions of the Zernike-based simulation was shown to be the case of a single $C_n^2$-slice at the halfway point of propagation means that we may replace the correlation kernel in these previous simulations with the correlation kernel in this paper. The result is maintaining the high degree of speed while dramatically improving the accuracy.

\subsubsection{Speed}
Comparing the speed of Zernike-based simulation and split-step leads us to consider what is a fair comparison. For the purposes of comparisons of speed, we consider only the time to actually generate the phase distortions, assuming all phase screens and integrals have been evaluated. In the case of split-step, this translates to the time to perform numerical wave propagation, while for multi-aperture we measure the time to generate the random fields using FFT-based generation and the approximation by Chimitt et al. \cite{Chimitt_2022}. The integral \eqref{eq: main_result_continuous2} may be computed \emph{once} offline at a very high resolution. This admittedly takes a good amount of computation. However, the result can then be stored and utilized repeatedly. Therefore, the time to compute the integral is unreasonable to include in such a comparison. Inclusion of the loading time for the integrals (and the subsequent evaluation of \eqref{eq: main_result_discrete}) could be included, however, this will be done once per configuration. One could still generate an infinite amount of data (all with different realizations) from one evaluation of \eqref{eq: main_result_discrete}. That being said, we still do include this timing measurement for completeness (though, the current operation is sequential with a sure possibility of being sped up through vectorized computation). This leaves the question as to whether or not phase screen generation time for split-step should be included. This would be unfair to a degree, as the same could be said for the phase screens which could be computed offline and stored. To make the comparison fair in this sense, we only consider the time it takes once everything has been precomputed. That being said, our method offers the benefit over phase screen generation due to the fact that the integral need only be evaluated once; the phase screens must be generated for every single independent run of split-step.



\begin{figure}
    \centering
    \includegraphics[width=0.75\linewidth]{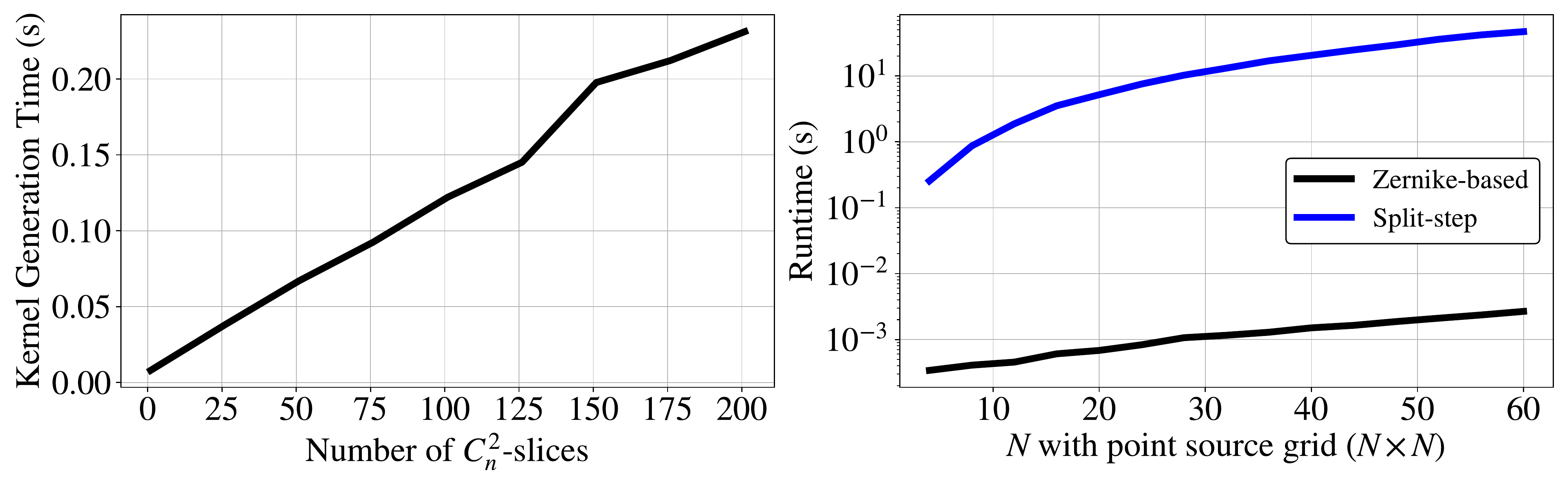}
    \caption{[Left] The time to load the precomputed integrals and form \eqref{eq: main_result_discrete} in the case of 36 Zernike polynomials is shown (utilizing approximation in Chimitt et al.\cite{Chimitt_2022}). Specifically, this is done for a grid of $64\times64$ points (representing the maximum case of the plot on the right.) [Right] The time comparison between our implementation between split-step and Zernike-based simulation for varying point source grids. This implementation of split-step uses 9 non-zero phase screens while the Zernike-based simulation again uses the approximation listed in Chimitt et al.\cite{Chimitt_2022} and 36 Zernike coefficients.}
    \label{fig: ma_vs_ss_speed}
\end{figure}

We feel this is the most fair comparison, as the two methods do not have a simple one-to-one correspondence of what they must initially generate. We present the results of this in \fref{fig: ma_vs_ss_speed}. We primarily have used the Python library `PyTorch' for both implementations as we have found in practice its FFT package to be more reliably faster than numpy. Note that we do not include PSF formation or application in this comparison. There are two reasons for this: (1) the methods both effectively generate phase realizations, thus the PSF generation and application would be identical, adding an equal upward shift to both methods (2) the series of publications which describes the multi-aperture method \cite{Chimitt2020, Mao_2021_ICCV, Chimitt_2022} utilizes engineering tricks to speed up exactly this process of PSF generation and application. The same tricks may not apply to split-step exactly, however, it is reasonable to expect some similar variety of these tricks may benefit split-step in terms of speed. Being that this is not the focus of this present paper, we do not consider them to be important parts of the comparison and thus do not include them.

\subsubsection{Resolution and Memory Issues}
Zernike-based simulations such as \cite{Chimitt2020, Mao_2021_ICCV} are harshly limited by the number of point sources they may simulate. This is due to the entire set of Zernike correlations being non-WSS as previously described. This large correlation matrix, and its decomposition, enforces the upper limit on the generation of high resolution spatial statistics. With $N_z$ Zernike coefficients and a grid of point sources of size $H\times W$, the required matrix will be of size $(N_z HW \times N_z HW)$. With high accuracy in the phase domain being a common requirement, $N_z$ will be large, significantly limiting the size on the spatial grid $H \times W$. 

\begin{figure}
    \centering
    \includegraphics[width=0.75\linewidth]{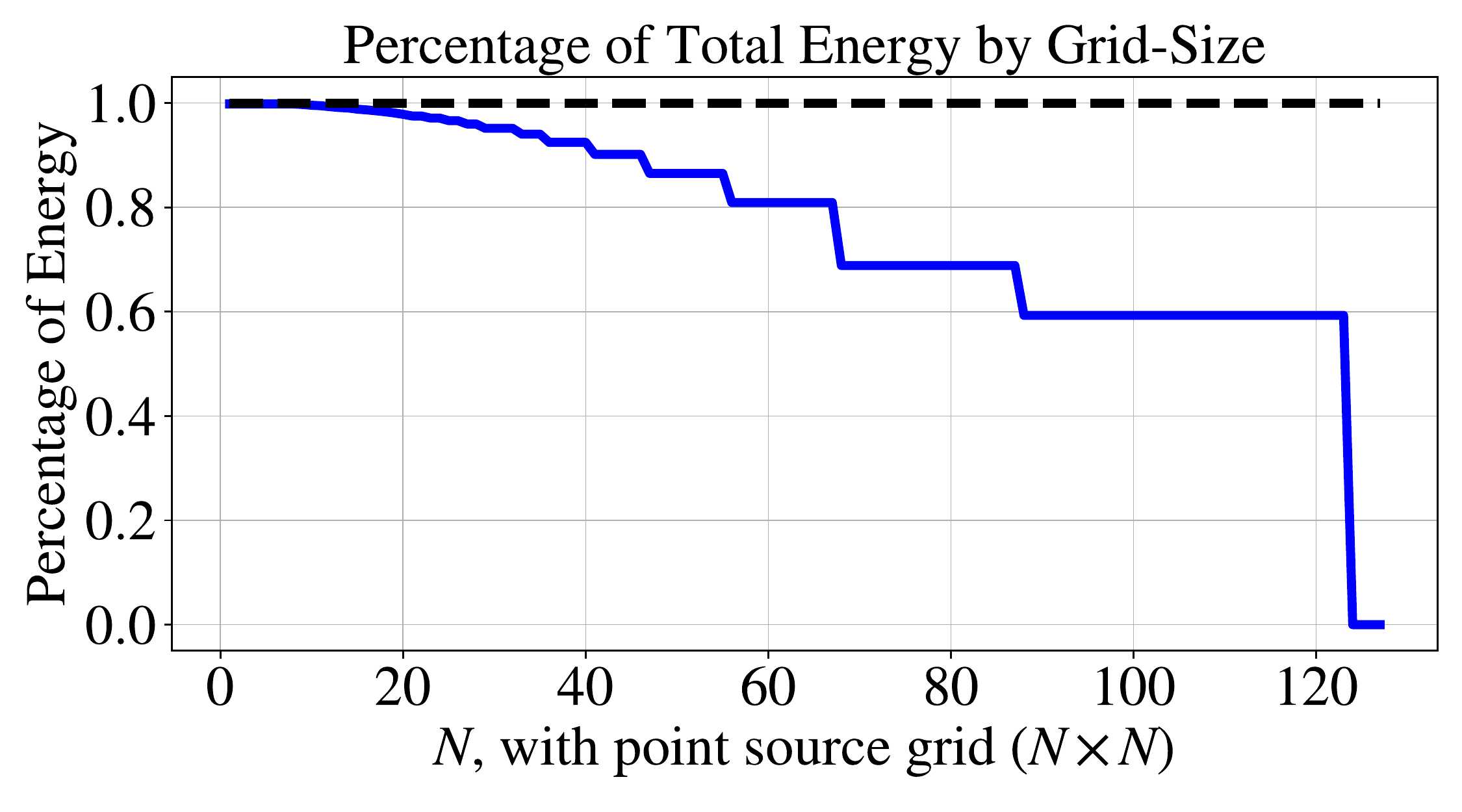}
    \caption{The proportion of total energy represented by the maximum allowable Zernike coefficients at the given grid size. The principle limitation here is the memory -- which is 16 GB in our case}.
    \label{fig: res_vs_acc}
\end{figure}

For the purposes of generating a single or independent Zernike coefficient fields, one may use FFT-based methods When extending to an image and multiple Zernike coefficients, this limit in sampling must be taken into account. To quantify the trade-off that exists within the current state of Zernike-based simulations for high-resolution images, we present the results in \fref{fig: res_vs_acc}, which shows the trade-off in accuracy that may be retained for an $N\times N$ grid of point sources. Accuracy here is measured relative to the total energy in the first 1275 Zernike coefficients. We then compare the energy of Zernike representations of $N_z$ coefficients to this total energy. The maximum resolution $N \times N$ is dictated by memory-limitations, in our case 16 GB; we set the maximum resolution according to a decomposition and generation that we are capable of performing on the described PC. This suggests Zernike-based simulations must rely on some additional approximation or a sampling methodology which does not require the covariance matrix to be used explicitly or instead does not rely on the WSS assumption. In the case of Mao et al. \cite{Mao_2021_ICCV}, a grid of $64\times 64$ with 36 Zernike coefficients each is then interpolated to the resolution of the image.

\section{Conclusion}
In this work, we've presented an alternative derivation for the correlations of Zernike coefficients for anisoplanatic turbulence. This enables previous versions of Zernike-based simulations to achieve a higher degree of accuracy while minimally compromising on speed. The correlations presented in this work have been shown to match the known angle-of-arrival correlations as well as explain some of the behavior in the mismatch of split-step's empirical correlations. Finally, we have outlined a problem facing Zernike-based simulations which restrict it from being directly used with FFT-based sampling.



\appendix
\section{Definition of Zernike Correlation Functions.}
\label{ap: takato}
Here we detail the function $f_{ij}$ which characterizes the correlations of the Zernike polynomials. The form of their equation is rather cumbersome and somewhat difficult to interpret for certain values of Noll indices. The purpose of this discussion is to simplify the resulting equation for the ease of interpretation and further study of these results. Following Takato and Yamaguchi \cite{Takato1995}, we first define the function
\begin{equation}
    I_{a, b, c} (s, k_0) = \int dx \frac{J_{a}(sx) J_{b}(x) J_{c}(x)}{x(x^2 + k_0)^2},
\end{equation}
with $J_k$ as the $k$th order Bessel function of the first kind. With Noll indices $(n_i, m_i) \to i$, we then define
\begin{align}
    n^+ &= n_i + n_j,\\
    n^- &= n_i - n_j,\\
    m^+ &= m_i + m_j,\\
    m^- &= m_i - m_j.
\end{align}
We also define an indicator function,
\begin{equation}
    h(i, j) = \begin{cases}
    1 & m_i \neq 0; m_j \neq 0; i+j \text{ even}\\
    2 & m_i \neq 0; m_j \neq 0; i+j \text{ odd}\\
    3 & m_i =0; j \text{ even} \oplus m_j = 0; i \text{ even}\\
    4 & m_i =0; j \text{ odd} \oplus m_j = 0; i \text{ odd}\\
    5 & m_i = 0; m_j = 0
    \end{cases},
\end{equation}
with $\oplus$ denoting the XOR function. First, we will present a form in line with that of \cite{Takato1995}, though using some of this notation and the appropriate simplifications. For a displacement $\vs=(s, \varphi)$ written in polar form, we can write the expression in \cite{Takato1995} as in \eqref{eq: takato_expression}
\vspace{6ex}
\begin{equation}
    f_{ij}(\vs, k_0) = 
    \begin{cases}
    \pm (-1)^{(n^+ - m^+)/2} \cos(m^+\varphi) I_{m^+, n_i + 1, n_j + 1}(2s, 2\pi R k_0) &\\
    \hspace{3ex}+ (-1)^{(n^+ +2m_i + |m^-|)/2} \cos(m^-\varphi) I_{|m^-|, n_i + 1, n_j + 1}(2s, 2\pi R k_0)& h(i,j) = 1\\
    (-1)^{(n^+ - m^+)/2} \sin(m^+\varphi) I_{m^+, n_i + 1, n_j + 1}(2s, 2\pi R k_0) &\\
    \hspace{3ex}+ (-1)^{(n^+ +2m_i + |m^-|)/2} \sin(m^-\varphi) I_{|m^-|, n_i + 1, n_j + 1}(2s, 2\pi R k_0) & h(i,j) = 2\\
    (-1)^{(n^+ - m^+)/2} \sqrt{2} \cos(m^+ \varphi) I_{m^+, n_i + 1, n_j + 1}(2s, 2\pi R k_0) & h(i,j) = 3\\
    (-1)^{(n^+ - m^+)/2} \sqrt{2} \sin(m^+ \varphi) I_{m^+, n_i + 1, n_j + 1}(2s, 2\pi R k_0) & h(i,j) = 4\\
    (-1)^{(n^+ - m^+)/2} I_{m^+, n_i + 1, n_j + 1}(2s, 2\pi R k_0) & h(i,j) = 5
    \end{cases},
    \label{eq: takato_expression}
\end{equation}
with the $\pm$ corresponding to $+$ if both $(i,j)$ are even, and $-$ if they are both odd.

However, our notation highlights further possible simplification. We can therefore write the function as
\begin{align}
    f_{ij}(\vs, k_0) = (-1)&^{(n^+ - m^+)/2}  \Theta^{(1)}(i,j) I_{m^+, n_i + 1, n_j + 1}(2s, 2\pi R k_0) \notag\nonumber\\
    + (-1)&^{(n^+ +2m_i + |m^-|)/2} \Theta^{(2)}(i,j) I_{|m^-|, n_i + 1, n_j + 1}(2s, 2\pi R k_0),
    \label{eq: takato_expression_us_final}
\end{align}
with functions
\begin{equation}
    \Theta^{(1)}(i,j) = \begin{cases}
    (-1)^{j} \cos(m^+ \varphi) & h(i,j) = 1 \\
    \sin(m^+ \varphi) & h(i,j) = 2 \\
    \sqrt{2}\cos(m^+ \varphi) & h(i,j) = 3 \\
    \sqrt{2}\sin(m^+ \varphi) & h(i,j) = 4 \\
    1 & h(i,j) = 5 \\
    \end{cases}
\end{equation} and,
\begin{equation}
    \Theta^{(2)}(i,j) = \begin{cases}
    \cos(m^- \varphi) & h(i,j) = 1 \\
    \sin(m^- \varphi) & h(i,j) = 2 \\
    0 & h(i,j) = 3 \\
    0 & h(i,j) = 4 \\
    0 & h(i,j) = 5 \\
    \end{cases},
\end{equation}
contributing the angular terms.

\subsection*{Funding}
The research is based upon work supported in part by the Intelligence Advanced Research Projects Activity (IARPA) under Contract No. 2022‐21102100004, and in part by the National Science Foundation under the grants CCSS-2030570 and IIS-2133032. The views and conclusions contained herein are those of the authors and should not be interpreted as necessarily representing the official policies, either expressed or implied, of IARPA, or the U.S. Government. The U.S. Government is authorized to reproduce and distribute reprints for governmental purposes notwithstanding any copyright annotation therein.

\bibliographystyle{plain}
\bibliography{references}

\end{document}